\documentstyle [12pt,twoside]{article}

\newcommand{\bq}{\begin{equation}}
\newcommand{\ba}{\begin{eqnarray}}
\newcommand{\eq}{\end{equation}}
\newcommand{\ea}{\end{eqnarray}}

\def\d{\delta}

\def\f{\phi}

\def\l{\lambda}

\def\D{\Delta}
\def\F{\Phi}
\def\G{\Gamma}

\def\bo{{\raise.15ex\hbox{\large$\Box$}}}
\def\bob{{\lower.2ex\hbox{\large$\Box$}}}
\def\pa{\partial}

\def\TH{{\raise.2ex\hbox{$\displaystyle \bigodot$}\mskip-4.7mu \llap H \;}}
\def\face{{\raise.2ex\hbox{$\displaystyle \bigodot$}\mskip-2.2mu \llap {$\ddot
        \smile$}}}

\def\Hat#1{\rlap{\kern.10em$\widehat{\phantom G}$}#1}
\def\HAt#1{\rlap{\kern.05em$\widehat{\phantom G}$}#1}

\def\cap#1{\rlap{\kern.1em$\widehat{\phantom{G\vrule height.8em}}$}#1{}}
\def\Cap#1{\rlap{\kern.05em$\widehat{\phantom{G\vrule height.8em}}$}#1{}}

\def\VEV#1{\left\langle #1\right\rangle}

\def\abs#1{\left| #1\right|}
\def\leftrightarrowfill{$\mathsurround=0pt \mathord\leftarrow \mkern-6mu
        \cleaders\hbox{$\mkern-2mu \mathord- \mkern-2mu$}\hfill
        \mkern-6mu \mathord\rightarrow$}
\def\overleftrightarrow#1{\vbox{\ialign{##\crcr
        \leftrightarrowfill\crcr\noalign{\kern-1pt\nointerlineskip}
        $\hfil\displaystyle{#1}\hfil$\crcr}}}

\def\frac#1#2{{\textstyle{#1\over\vphantom2\smash{\raise.20ex
        \hbox{$\scriptstyle{#2}$}}}}}

\catcode`@=11
\def\underline#1{\relax\ifmmode\@@underline#1\else
        $\@@underline{\hbox{#1}}$\relax\fi}
\catcode`@=12

\def\nis{\nointerlineskip}
\def\Abar{\vbox{\nis\moveright.33em\vbox{
        \hrule width.35em height.04em}\nis\kern.05em\hbox{$A$}}{}}
\def\Dbar{\vbox{\nis\moveright.20em\vbox{
        \hrule width.50em height.04em}\nis\kern.05em\hbox{$D$}}{}}
\def\Gbar{\vbox{\nis\moveright.20em\vbox{
        \hrule width.50em height.04em}\nis\kern.05em\hbox{$G$}}{}}
\def\mbar{\vbox{\nis\moveright.15em\vbox{
        \hrule width.60em height.04em}\nis\kern.05em\hbox{$m$}}{}}
\def\Rbar{\vbox{\nis\moveright.20em\vbox{
        \hrule width.50em height.04em}\nis\kern.05em\hbox{$R$}}{}}
\def\Vbar{\vbox{\nis\moveright.05em\vbox{
        \hrule width.60em height.04em}\nis\kern.05em\hbox{$V$}}{}}
\def\Xbar{\vbox{\nis\moveright.20em\vbox{
        \hrule width.60em height.04em}\nis\kern.05em\hbox{$X$}}{}}
\def\thetabar{\vbox{\nis\moveright.15em\vbox{
        \hrule width.30em height.04em}\nis\kern.05em\hbox{$\theta$}}{}}
\def\Lambdabar{\vbox{\nis\moveright.25em\vbox{
        \hrule width.35em height.04em}\nis\kern.05em\hbox{${\mit\Lambda}$}}{}}
\def\Sigmabar{\vbox{\nis\moveright.25em\vbox{
        \hrule width.50em height.04em}\nis\kern.05em\hbox{${\mit\Sigma}$}}{}}
\def\phibar{\vbox{\nis\moveright.18em\vbox{
        \hrule width.40em height.04em}\nis\kern.05em\hbox{$\phi$}}{}}
\def\chibar{\vbox{\nis\moveright.12em\vbox{
        \hrule width.40em height.04em}\nis\kern.05em\hbox{$\chi$}}{}}
\def\psibar{\vbox{\nis\moveright.23em\vbox{
        \hrule width.40em height.04em}\nis\kern.05em\hbox{$\psi$}}{}}
\def\debar{\vbox{\nis\moveright.18em\vbox{
        \hrule width.35em height.04em}\nis\kern.05em\hbox{$\partial$}}{}}
\def\delbar{\vbox{\nis\moveright.10em\vbox{
        \hrule width.63em height.04em}\nis\kern.05em\hbox{$\nabla$}}{}}

\begin{document}

\centerline{\bf Multiplicative Noise: Applications in Cosmology and
Field Theory}

\vspace{.5cm}

\centerline{Salman Habib$^{\dagger}$}

\vspace{.2cm}

\centerline{\em{T-6, Theoretical Division}}
\centerline{\em{Los Alamos National Laboratory}}
\centerline{\em{Los Alamos, NM 87545}}
\centerline{\em{U.S.A.}}

\vspace{1.5cm}

\centerline{To appear in {\em Stochastic Processes in Astrophysics}}
\centerline{Proceedings of the Eighth Annual Workshop in Nonlinear
Astronomy}
\centerline{Gainesville, Florida, 4 - 6 February, 1993}

\vspace{1.5cm}

\centerline{\bf Abstract}

Physical situations involving multiplicative noise arise generically
in cosmology and field theory. In this paper, the focus is first on
exact nonlinear Langevin equations, appropriate in a cosmological
setting, for a system with one degree of freedom. The Langevin
equations are derived using an appropriate time-dependent
generalization of a model due to Zwanzig. These models are then
extended to field theories and the generation of multiplicative noise
in such a context is discussed. Important issues in both the
cosmological and field theoretic cases are the fluctuation-dissipation
relations and the relaxation time scale. Of some importance in
cosmology is the fact that multiplicative noise can substantially
reduce the relaxation time. In the field theoretic context such a
noise can lead to a significant enhancement in the nucleation rate of
topological defects.

\vfill \noindent $^{\dagger}$e-mail: habib@eagle.lanl.gov
\newpage

\centerline{\bf I. Introduction}

The study of noise in a field theoretic context has acquired ever
increasing importance in the last few decades. More or less
traditional applications such as phenomenological theories of spinodal
decomposition have been augmented recently by several new research
areas, e.g., stochastic quantization \cite{PW}, noise-induced
transitions \cite{HL}, stochastic inflation \cite{AAS}, and quantum
decoherence \cite{QD}. The last two applications are of interest in
cosmology.

The objective here is to concentrate on the behavior of systems
interacting with multiplicative noise in a few examples and to
contrast it with what happens when the noise is additive. In the
cosmological situation one also has to deal with the complication
that, since the Universe is expanding, there is a background time
dependence of various parameters that would otherwise be fixed.

Over the past two decades or so, the problem of understanding
statistical field theory in a cosmological context has received much
attention. One of the key issues here is to allow correctly for the
time dependence of the background spacetime. Apart from any purely
abstract motivation, this problem is of great practical interest as far
as understanding the physics of the early Universe is concerned. Since
not much is known about the early history of the Universe,
approximations are often made which, although not justified
rigorously, seem physically well motivated and even essential, if one
wishes to obtain concrete results.

A standard assumption is the idea that, at some level, the physical
degrees of freedom of the Universe can be split into two coupled
pieces, a ``system'' component, the evolution of which is to be
followed in some detail, and a ``bath'' component, the detailed
evolution of which is taken to be irrelevant. Examples of this picture
include: (1) An understanding of inflation in terms of an inflaton
field evolving in the rest of the Universe, which serves as an
external environment \cite{RB}-\cite{HKB}, (2) the general
notion of ``coarse-graining'' as a physical mechanism in terms of
which to extract quantum decoherence, thus leading to a
``quantum-to-classical'' transition in the early Universe
\cite{DECO}, (3) the systematic development of statistical quantum
field theory \cite{BLH}, using a closed-time-path formalism to
derive quantum dissipation and memory loss, (4) the intriguing, but
not yet completely understood \cite{SHSI}, program of stochastic
inflation originally proposed by Starobinsky \cite{AAS}.

Much of the analysis has proceeded making use of essentially {\em ad
hoc} Fokker-Planck equations, with an assumed bilinear interaction
between the system and the bath. It is important to understand in what
sense these models are physically reasonable, and, how relaxing the basic
assumption of bilinearity changes the underlying physics, i.e., what
happens when the noise in these models is multiplicative?

To understand some of these issues, it is instructive to consider
models of systems interacting with baths consisting of time-dependent
harmonic oscillators. In these models it is possible to derive exact,
nonlocal Langevin equations \cite{HKB}. These models are
nonlinear, time-dependent generalizations of phenomenological models
which have proven quite successful in other branches of physics, and,
{\em are} well motivated phenomenologically, even if they are not
derivable strictly from first principles.

The analysis of Habib and Kandrup \cite{HKB} showed that (1) allowing
for a nontrivial time-dependence necessarily induces qualitatively new
effects like a mass (or frequency) ``renormalization,'' even for the
special case of bilinear couplings \cite{HKA}; and, moreover, (2)
allowing for nonlinearities in the system-environment coupling induces
new effects aside from the usual ``friction'' term. For example,
nonlinearities give rise to an additional renormalization of the
system potential, and they lead to a ``memory'' kernel which involves
the state of the system. In the presence of nonlinearities, one's
naive intuition is not completely lost: Even allowing for
nonlinearities and time-dependent couplings in the interaction between
the system and environment, for the case of time-independent
oscillators (where the bath may still be viewed as being ``at
equilibrium'') it is possible to derive a simple
fluctuation-dissipation theorem \cite{RWZ}. However, when the
oscillators become time-dependent, the fluctuation-dissipation theorem
is no longer exact. Nevertheless, the theorem is approximately valid
as long as the coupling between the system and environment is dominated by
modes of sufficiently short wavelength. In a cosmological setting this
implies that, on scales short compared with the horizon length, it is
still possible to speak of an approximate equilibrium and an approach
towards that equilibrium.

The examples treated here that are relevant in a field theoretic
context refer to stochastic equations for such systems where the
noise is a spacetime noise rather than the Gaussian, white noise
encountered usually. By discretizing in space, it is possible in these
cases to use oscillator models for the heat bath and to derive exact
Langevin equations. Since these equations are at the heart of one
approach to finite temperature simulations of field theories, it is
useful to have a simple (indeed trivial) derivation for them. The two
specific cases considered here will be a double-well $\f^4$ theory in
$1+1$-dimensions and the nonlinear sigma model in $1+1$-dimensions.
The last example is an interesting one: even though the coupling to
the heat bath is bilinear, the noise is multiplicative. This happens
because the stochastic forcing term must be consistent with a
constraint.

Section II will be devoted to the oscillator model and its
generalizations. The equations of motion will be derived and the
conditions under which a Langevin description arises will be
discussed. Particular attention will focus on fluctuation-dissipation
relations, time dependent baths, multiple degree of freedom systems,
and nonlinear system-bath couplings. Section III will deal with the
modification of the relaxation time scale by multiplicative noise. The
method of Lindenberg and Seshadri \cite{LS} (first suggested by
Stratonovich \cite{RLS}) will be applied to a cosmological model
problem. It will be shown that substantial changes in the relaxation
rate can occur. Two examples studied recently provide numerical
evidence: a harmonic oscillator driven to equilibrium by quadratic
multiplicative noise and the nucleation of kinks in the double-well
$\f^4$ theory in $1+1$-dimensions. In the first case, the enhancement
of the relaxation rate predicted by the theory is consistent with the
simulations, while in the second case an enhancement of the kink
nucleation rate is observed (this process has not yet been
investigated analytically). Finally, Section IV will offer some
conclusions, unanswered questions, and speculations regarding future
results.

\newpage

\centerline{\bf II. Multiplicative Noise, Time Dependence, and the
Oscillator Heat Bath}

The material at the beginning of this section borrows heavily from
Ref. \cite{HKB} where the reader will find a rather more detailed
discussion of the derivation presented here.

The idea behind the oscillator heat bath model is relatively simple:
given some arbitrary composite Hamiltonian one splits it into three
pieces: the system, the bath, and the interaction. If this formal
split is to be useful, the system must be ``small'' in some sense when
compared to the bath or environment. As a practical matter what this
means is that, as far as the system is concerned, the full Hamiltonian
\bq
H_T=H_S+H_B+H_I \label{1}
\eq
is well approximated by
\bq
H=H_S+\delta H_B+\delta H_I \label{2}
\eq
where $\delta H_B$ is the Hamiltonian for a collection of harmonic
oscillators and $\delta H_I$ an interaction Hamiltonian linear in the
oscillator variables $q_A$ \cite{MS} \cite{CLA}. Basically,
one assumes that, in the absence of any coupling with the system, the
environment is characterized by some fixed, possibly time-dependent,
solution. Allowing for a weak coupling with the system (weak in the
sense that each bath mode is only changed marginally), one then
identifies the $q_A$'s as perturbed variables, i.e., degrees of
freedom defined relative to the background solution (e.g., phonons).
This has two implications: (1) the environment can be visualized as a
collection of oscillators with (possibly time-dependent) frequencies,
so that ${\delta}H_{B}$ is quadratic in bath variables $q_A$, and, (2)
because the interaction of the environment with the system is assumed
to be weak, ${\delta}H_{I}$ must be linear in the $q_{A}$'s. Note that
the system is not necessarily weakly altered, so the interaction
${\delta}H_{I}$ does not have to be linear in the system variable $x$.
In principle one can proceed without imposing any restrictions on the
form of the system Hamiltonian $H_{S}$.

With the above set of assumptions, one can write
\bq
H_{S}={1\over 2}{\;}v^{2}+V_{ren}(x,t), \label{3}
\eq
\bq
\delta H_{B}={1\over
2}\sum_{A}\left[p_{A}^{~2}+{\Omega}_{A}^{~2}(t)q_{A}^{~2}\right],
\label{4}
\eq
and, in terms of relatively arbitrary functions $\Gamma_A$,
\bq
\delta H_{I}=-{\,} \sum_{A}{\,} \Omega_A^{~2}(t)\Gamma_{A}(x,t)q_{A}.
\label{5}
\eq
It is, however, convenient to rewrite $H$ in the manifestly positive
form
\bq
H={1\over 2}v^{2}+U(x,t) +
{1\over 2}\sum_{A}\left\{p_{A}^{~2}+{\Omega}_{A}^{~2}(t)
\left[q_{A}-{\Gamma}_{A}(x,t)\right]^2\right\}, \label{6}
\eq
where in terms of the ``renormalized'' potential $V_{ren}$,
\bq
U(x,t)=V_{ren}-{1\over 2}\sum_A\Omega_A^{~2}(t)\Gamma_A^{~2}(x,t).
\label{7}
\eq

Couplings of a system to some environment can induce finite and
stochastic renormalizations in the system potential, although this is
not always so \cite{CLA}. In this paper, the words ``system
potential'' will always refer explicitly to the renormalized system
potential.

Finally, as emphasized, e.g., by Caldeira and Leggett \cite{CLA},
it should be stressed that this is more than simply a toy model.
This form of the Hamiltonian generally provides a correct description
for {\em any} system which is only weakly coupled to its surroundings
(Although there are caveats relating to the spectral distribution of
the modes: one has to be sure that there are no negative frequency,
unstable modes. This is not a trivial issue when the full Hamiltonian
is complicated and not well understood). To facilitate a concrete
calculation, this Hamiltonian need only be supplemented by two inputs,
namely the spectral distribution of the environmental modes and the
form of the coupling to the system. For a general physical problem,
these may either be extracted from experimental data or derived from
theoretical considerations (of course, this may not turn out to be
simple in practice).

The equations of motion generated from the Hamiltonian (\ref{6})
clearly take the forms
\ba
\dot{x}&=&v, \nonumber\\
\dot{v}&=&-{\pa\over\pa x}{\,}U(x,t)
+\sum_A {\Omega}_{A}^{~2}(t)\left[q_A-\Gamma_A(x,t)\right]
{\pa\over\pa x}\G_A(x,t), \nonumber\\
\dot{q}_A&=&p_A, \nonumber\\
\dot{p}_A&=&-{\Omega}_{A}^{~2}
\left[q_A-\Gamma_A(x,t)\right], \label{8a}
\ea
where an overdot denotes a time derivative ${\partial}/{\partial}t$.
Following Ref. \cite{HKB} one can show that these equations collapse
to the single nonlocal equation
\bq
\dot{v}=-{{\partial}U\over {\partial}x}
-\int_0^{t}ds \left[K(t,s)v(s)+M(t,s)x(s)\right] + F_s(t), \label{16}
\eq
where, in terms of the quantities
\bq
A_A(s,t)=\left(\sum_m \gamma_{A}^{(m)}(s)x^{m-1}(s)\right)
         \left(\sum_n \gamma_{A}^{(n)}(t)x^{n-1}(t)\right) \label{17}
\eq
and
\bq
B_A(s,t)={1\over n}{{\partial}\over {\partial}s}A_A(t,s), \label{18}
\eq
the ``memory kernels'' $K(t,s)$ and $M(t,s)$ are
\ba
K(t,s)&=&\sum_A\Omega_A^{~2}(t)A_A(s,t)W_A(s,t), \label{19}\\
M(t,s)&=&\sum_A\Omega_A^{~2}(t)B_A(s,t)W_A(s,t).  \label{20}
\ea
Here one has assumed a polynomial system-bath coupling
\bq
{\Gamma}_{A}(x,t)= \sum_{n=1}^{N} {1\over n}\gamma_{A}^{(n)}(t)x^{n},
\label{15}
\eq
and the force $F_s$ is
\bq
F_s(t)=\sum_A {\Omega}_{A}^{~2}(t)
\sum_m \gamma_{A}^{(m)}(t)x^{m-1}
\left\{
\left[q_A(0)-\Gamma_{A}(x,0)\right]C_A(t)+p_A(0)S_A(t)\right\},
\label{21}
\eq
where $S_A(t)$ and $C_A(t)$ denote two linearly independent solutions
to the homogeneous oscillator equation
\bq
\ddot{\Xi}_A+{\Omega}_{A}^{~2}(t){\Xi}_A=0,  \label{9}
\eq
satisfying the initial conditions $C_A(0)=\dot{S}_A(0)=1$ and
$S_A(0)=\dot{C}_A(0)=0$.

Equation (\ref{16}) does not at first sight appear to be an ordinary
Langevin equation. As a first step to see how this may be so,
it is possible to prove a generalized fluctuation-dissipation theorem,
even if $\Gamma_A$ is a nonlinear function of $x$ and/or explicitly
time-dependent \cite {HKB}(as long as the oscillator frequencies
${\Omega}_{A}$ are not time-dependent).

Consider an ensemble of initial conditions for which the first moments
vanish identically, i.e.,
\bq
\VEV{Q_A(0)}=\VEV{{p_A}(0)}\equiv 0, \label{22}
\eq
with $Q_A(0)\equiv q_A(0)-\Gamma_A(x(0),0)$, and where the second
moments are initially thermal, so that
\bq
\VEV{{p_A(0)p_B(0)}}=
{\Omega}_{A}(0){\Omega}_{B}(0)\VEV{Q_A(0)Q_B(0)}
=k_BT\delta_{AB}, \label{23}
\eq
where the angular brackets denote an initial ensemble average. Then
\bq
\VEV{F_s(t)}=0 \label{24a}
\eq
and
\bq
\VEV{F_s(t_1)F_s(t_2)}=k_BTK(t_1,t_2), \label{24}
\eq
thereby identifying $F_s(t)$ as a noise and proving a generalized
fluctuation-dissipation theorem linking the noise autocorrelator with
the ``viscosity kernel'' $K(t,s)$. Equation (\ref{16}) can now be
viewed as a nonlinear, nonlocal Langevin equation. It is important to
note that for arbitrary time dependence of the oscillator frequencies,
the fluctuation-dissipation theorem does not hold. In the cosmological
context, an approximate fluctuation-dissipation relation is valid as
long as the system time scale is small compared to the expansion time
scale of the Universe.

The Langevin equation (\ref{16}) reduces to a well known form in an
appropriate limit. If one neglects nonlinearities in the coupling
between system and bath, assuming that $\Gamma_A\propto x$, one
immediately recovers a special model considered in Ref. \cite{HKA}.
With the further assumption that ${\Omega}_{A}$ and $\Gamma_{A}$ are
independent of time, one is reduced to the well known independent
oscillator (IO) model \cite{IO}. It is thus possible to address
systematically the question of how the incorporation of nonlinearities
and/or time-dependences leads to systematic changes in the Langevin
equation derived for that original model. Ref. \cite{HKB} discusses
such issues as time scale separations and Markov limits (when the
nonlocal in time equation becomes local). Here it will be simply
assumed that it is indeed possible to write (\ref{16}) as a local,
albeit nonlinear Langevin equation. Once this can be done, it is
immediate to write the Fokker-Planck equation (for the time evolution
of the phase space distribution function) that corresponds to a
particular Langevin equation.

Can this simple approach be extended to continuous systems, e.g.,
field theories? When discretized, classical field theories can be
thought of as interacting continuous spin systems. In this case, the
model works more or less as usual except that the system has a very
large number of degrees of freedom and the effective noise turns out
to be a ``spacetime noise'' rather than the one considered so far. The
Langevin equations derived in this way are useful for thermal
simulations as an alternative approach to Monte Carlo techniques. In
what follows no explicit time dependences will be assumed.

As a relevant example of a field theoretic problem, a double-well
$\l\f^4$ theory in $1+1$-dimensions is particularly convenient. This
theory admits stationary wave solutions (``kinks'') and the
statistical mechanics of these objects has recently been carefully
investigated via Langevin simulations \cite{FJSH}.

The Lagrangian for this theory is
\bq
L={1\over 2}(\pa_t\F)^2-{1\over 2}(\pa_x\F)^2+{1\over
2}m^2\F^2-{1\over 4}\l\F^4.                      \label{25}
\eq

In order to derive the appropriate Langevin equation for this theory
it is convenient to discretize in space and use the Hamiltonian
formulation. The discrete Hamiltonian
\bq
H_D=\sum_i {1\over 2}\left[p_i^{~2}-m^2\F^2_{i}+{1\over
2}\l\F^4_{i}+{1\over 2}\left({{\F_{i+1}-\F_i}\over\D}\right)^2 +
{1\over 2}\left({{\F_{i}-\F_{i-1}}\over\D}\right)^2\right] \label{26}
\eq
gives rise to the equations of motion
\bq
\pa_{tt}^2\F_i=\D_D\F_i+\F_i(1-\F_i^{~2})
\eq
where $\D$ is the lattice spacing and the spatial lattice Laplacian
\bq
\D_D\F_i={{\F_{i+1}+\F_{i-1}-2\F_i}\over 2 \D^2}.
\eq
It is clear that $H_D$ can be interpreted as describing a continuous
spin system with spins $\F_i$ (at the lattice site $i$) interacting
via nearest neighbor interactions. To drive this system to equilibrium
the idea is now to couple each spin to an independent oscillator heat
bath: there will now be as many ``baths'' as there are spins, all with
the same initial conditions in the sense that the averages (\ref{22})
and (\ref{23}) hold for each collection of oscillators. By following
the derivation given above, but for each individual spin this time, it
is easy to derive the Langevin equation (written, for simplicity, for
a linear coupling to the heat bath)
\bq
\pa_{tt}^2\F_i=\D_D\F_i-\eta\pa_t\F_i+\F_i(1-\F_i^{~2})+F_i \label{28}
\eq
where the noise auto-correlation
\bq
\VEV{F_i(t)F_j(s)}=2\eta k_BT\d(t-s)(\d_{ij}/\D).
\eq
Note that by taking independent heat baths for the spins we have
enforced the vanishing of the noise cross-correlation. Since the
fluctuation-dissipation relation is still valid, we know that that the
system of Langevin equations (\ref{28}) will drive the system to
thermal equilibrium. The continuum limit of (\ref{28}) is the Langevin
equation
\bq
\pa_{tt}^2\F=\pa_{xx}^2\F-\eta\pa_t\F+\F(1-\F^2)+F(x,t)
\eq
and in this limit the noise correlator
\bq
\VEV{F(x,t)F(y,s)}=2\eta k_BT\d(t-s)\d(x-y).
\eq
describes a white, ``spacetime'' noise.

One way of introducing multiplicative noise is
through the imposition of constraints on the system variables. One
enforces the noise to act in such a way that the reduced dynamics for
the system respects the constraint imposed via a Lagrange multiplier.
This process turns what would have been additive noise into
multiplicative noise: even if the coupling to the environment is
linear in the Hamiltonian, the effective Langevin equation contains
multiplicative noise.

As a representative example, consider the $O(3)$ nonlinear sigma model
in $1+1$-dimensions. The full Hamiltonian is
\bq
H=\int dx~{1\over 2}\left(\hat{p}\cdot\hat{p}+
\pa_x\hat{n}\cdot\pa_x\hat{n}\right),
\eq
with the constraint,
\bq
\hat{n}\cdot\hat{n}=1.
\eq
To implement the constraint, one introduces a Lagrange multiplier
$\l$ and writes the Hamiltonian as
\bq
H=\int dx~{1\over 2}\left(\hat{p}\cdot\hat{p}+
\pa_x\hat{n}\cdot\pa_x\hat{n}+\l(\hat{n}\cdot\hat{n}-1)\right).
\eq
The equation of motion is then
\bq
(-\pa_{tt}^2+\pa_{xx}^2)\hat{n}+\l\hat{n}=0
\eq
The Lagrange multiplier can be eliminated from this equation by taking
the dot product of it with $\hat{n}$. This procedure yields
\bq
(-\pa_{tt}^2+\pa_{xx}^2)\hat{n}-(\hat{n}\cdot(-\pa_{tt}^2+
\pa_{xx}^2)\hat{n})\hat{n}=0.
\eq
One can now discretize the Hamiltonian in much the same way as the
$\f^4$ theory considered above. Following that analysis one linearly
couples oscillator heat baths to each spin component at each lattice
point and derives in the usual way, the Langevin equation
\bq
\pa_{tt}^2\hat{n}-\pa_{xx}^2\hat{n}-\l\hat{n}+
\eta\pa_t\hat{n}+\hat{F}=0
\eq
Again, by eliminating the Lagrange multiplier, one obtains
\bq
\pa_{tt}^2\hat{n}-\pa_{xx}^2\hat{n}+(\hat{n}\cdot\pa_{xx}^2
\hat{n})\hat{n}+(\pa_t\hat{n}\cdot\pa_t\hat{n})\hat{n}+
\eta(\pa_t\hat{n}-(\hat{n}\cdot\pa_t\hat{n})\hat{n})+\hat{F}-
(\hat{n}\cdot\hat{F})\hat{n}=0,
\eq
a nonlinear Langevin equation with multiplicative noise. This equation
(supplemented with a symmetry breaking term) has been used in a
numerical study of thermal sphalerons in the nonlinear sigma model
\cite{NTS} where it has been verified that it takes the system to the
correct equilibrium distribution while maintaining the constraint to a
reasonable level of accuracy ($1$ part in $10^4$, the error due to
time-stepping inaccuracies). This way of driving the sigma model to
equilibrium is better than using variables which explicitly maintain
the constraint (for example, the polar angles) since there are no
coordinate singularities to cause dynamic range problems in
numerical simulations.

\newpage
\centerline{\bf III. Relaxation with Multiplicative Noise}

In the previous section multiplicative noise entered in the IO model
in two ways, first through a nonlinear coupling to the environment and
second, through the imposition of a constraint on the system
variables (which effectively introduces a nonlinear coupling). Systems
subjected to noise of this kind can have striking departures in their
dynamical behavior when compared to the case of additive noise. Here
the focus will be on the process of relaxation to equilibrium: Even
though multiplicative noise takes the system to the usual thermal
equilibrium state (when the fluctuation-dissipation theorem is valid),
the relaxation time scale can be quite different from that for
additive noise. In particular, it can be much smaller.

As an illustrative example, consider the cosmologically relevant IO
model of a ``system'' variable interacting with scalar ``radiation''
in a spatially flat, Friedmann-Robertson-Walker universe (discussed
more fully in Ref. \cite{HKB}, from where this example is taken)
\bq
H={1\over 2}v^{2}+U(x,{\eta}) +
{1\over 2}\sum_{A}\left\{ p_{A}^{~2}+
{\Omega}_{A}^{~2}({\eta})\left[q_{A}-\Gamma_{A}(x,
{\eta})\right]\right\}^{2}, \label{50}
\eq
where, in terms of constants $\lambda$, $\theta$, and $\sigma$,
\bq
U(x,{\eta})={1\over 4}\l ({\eta})x^{4}+
{1\over 3}{\theta}({\eta}) x^{3} +
{1\over 2}{\sigma}({\eta}) x^{2}, \label{51}
\eq
and the coupling
\bq
\Gamma_{A}(x,{\eta})={\gamma}^{(1)}_{A}({\eta})x +
{\gamma}^{(2)}_{A}({\eta})x^{2}. \label{52}
\eq
The frequencies ${\Omega}_{A}$ satisfy
\bq
{\Omega}_{A}^{~2}({\eta})={\omega}_{A}^{~2}-(1-6{\xi}){a''\over a}
\label{53}
\eq
where a prime $'$ denotes a conformal time derivative
${\partial}/{\partial}{\eta}$ and $a$ is the scale factor. The
conformal time $\eta$ is related to the cosmic time $t$ via
$d\eta=a^{-1}dt$.

If the scale factor $a$ has a simple power law time dependence
$a=t^{p}$, with $p\geq 1/2$, it follows that
\bq
a{\,}{\propto}{\,}({\eta}-{\eta}_{0})^{p/(1-p)}, \label{54}
\eq
where ${\eta}_{0}$ denotes an integration constant, so that
\bq
{\Omega}_{A}^{~2}({\eta})={\omega}_{A}^{~2}-
{\nu^{2}\over ({\eta}-{\eta}_{0})^{2}}, \label{55}
\eq
where
\bq
\nu^{2}=(1-6{\xi}){p(2p-1)\over (1-p)^{2}} \label{56}
\eq
is intrinsically positive when $p>1/2$ and $1-6\xi \geq 0$.
The quantities ${\omega}_{A}^{~2}$ denote eigenvalues of the spatial
Laplacian.

To proceed further, one obviously needs to determine the mode functions
${\Xi}_{A}$ for the time-dependent frequencies. The equation
\bq
{\Xi}_{A}'' +\left[{\omega}_{A}^{~2} - {\nu^{2}\over
\left({\eta}-{\eta}_{0}\right)^{2}}\right]{\Xi}_{A} = 0 \label{57}
\eq
is solved by a general
\bq
{\Xi}_{A}=\left[{\omega}_{A}\left({\eta}-
{\eta}_{0}\right)\right]^{1/2}
Z_{\mu}\left({\omega}_{A}\left({\eta}-{\eta}_{0}\right)\right),
\label{58}
\eq
where $Z_{\mu}$ denotes an arbitrary solution to Bessel's equation of
order
\bq
{\mu}^{2}=(1-6{\xi}){\,}{p(2p-1)\over (1-p)^{2}} + {1\over 4}=
{\nu}^{2}+ {1\over 4}, \label{59}
\eq

For the special cases ${\xi}=1/6$ (conformal coupling)
and/or  $p=1/2$ (a Universe dominated by electromagnetic
radiation), ${\mu}=1/2$ and the solutions ${\Xi}_{A}$ reduce to sines
and cosines. For these particular values, the bath Hamiltonian
${\delta}H_{B}$ is {\em time-independent} in the conformal frame. This
implies that the physical frequencies are simply red-shifted uniformly
as the Universe expands.

Even when the special conditions given above do not hold, as long as
the system time scale is small compared to the expansion time scale,
one recovers a relatively simple nonlocal Langevin equation, for which
the only explicit time dependences are in the potential $U$ and the
couplings between the system and the environment. It thus follows
that, in this approximation, a fluctuation-dissipation theorem holds,
so that one would anticipate an evolution towards some steady state
solution at late times. The fluctuation-dissipation theorem is of
course exact when $\xi=1/6$ and/or $p=1/2$.

If a Markov approximation is valid, the nonlocal Langevin equation can
be well approximated by a local equation of the form
\bq
v'=-{{\partial}U\over {\partial}x} -
{\cal M}(\eta)x(\eta) - {\cal K}(\eta)v(\eta) + F_{s}(\eta), \label{60}
\eq
where, for an initial thermal ensemble,
\bq
{\langle}F_{s}(\eta)F_{s}(s){\rangle}=
2k_{B}T{\cal K}(x,{\eta}){\delta}_{D}(\eta-s) \label{61}
\eq
This local Langevin equation leads immediately to a Fokker-Planck
equation of the form
\bq
{{\partial}f\over {\partial}{\eta}} +
{{\partial}\over {\partial}x}(vf) +
{{\partial}\over {\partial}v}\left[\left(-{\pa U\over\pa x} - {\cal
M}x - {\cal K}v\right)f\right] - k_{B}T{\cal
K}{\,}{{\partial}^{2}f\over {\partial}v^{2}}=0. \label{62}
\eq

Translated back into the physical frame, the Langevin equation
(\ref{60}) becomes
\bq
\dot{V}=-{1\over a^4}{\pa U\over\pa X}-{\bf
M}(t)X-\left[3{\dot{a}\over a}+{\bf K}(t)\right]V+{\bf F}_s(t), \label{63}
\eq
where the overdot denotes differentiation with respect to cosmic
time, and
\ba
X&=&{x\over a},               \label{a1}\\
V&=&\dot{X}={v\over a^2}-{\dot{a}\over a^2}x, \label{a2}\\
{\bf K}(t)&=&{{\cal K}\over a},               \label{a3}\\
{\bf M}(t)&=&{{\cal M}\over a^2}+{\bf K}(t)\left({\dot{a}\over
a}\right)+
\left({\dot{a}\over a}\right)^2+{\ddot{a}\over a},
\label{a4}\\
{\bf F}_s(t)&=&{F_s\over a^3}.  \label{a5}
\ea
The noise autocorrelator is now
\bq
\VEV{{\bf F}_s(t){\bf F}_s(t')}=2{k_BT\over a^4}{\,}{\bf K}(t)\d(t-t').
\label{a6}
\eq
Note that, in the physical frame, there are two sources of
damping, namely the viscosity $\propto {\bf K}V$ and the
cosmological frame-dragging $\propto H\equiv\dot{a}/a$.
The corresponding Fokker-Planck equation is now
\bq
{\pa f_p\over\pa t}+{\pa\over\pa X}(Vf_p)+{\pa\over\pa
V}\left\{\left[-{1\over a^4}{\pa U\over\pa X}-{\bf
M}(t)X-\left(3{\dot{a}\over a}+{\bf
K}(t)\right)V\right]f_p\right\}-{k_BT\over a^4}{\bf
K}(t){\pa^2f_p\over\pa V^2}=0.      \label{a7}
\eq
Note that the conformal temperature $T$ is related to the physical
temperature $T_{ph}$ by $T_{ph}=T/a$.

The analysis of the approach to equilibrium now follows that of
Lindenberg and Seshadri \cite{LS} based on earlier work by
Stratonovich \cite{RLS}. The basic idea is that for sufficiently weak
damping the time scale on which the system coordinate changes is much
faster than the time scale on which the energy evolves. This allows
one to average over the fast coordinates and convert the original
Langevin equation, written in its Fokker-Planck representation, to a
one-variable Fokker-Planck equation for the energy distribution
function (hence the name, ``energy-envelope method'').

As a first step one implements a change of variables from $(x,v)$ to
$(x,E)$, where, $E={1\over 2}v^{2}+{\cal U}(x,{\eta})$, to obtain a
new Fokker-Planck equation for the distribution $W(x,E,t)$ satisfying
\bq
f(x,v,\eta)dxdv=W(x,E,\eta)dxdE. \label{64}
\eq
To the extent that the energy is approximately conserved during a
single oscillation of the system, one can assume that
\bq
W(x,E,\eta)=\{2{\Phi}'(E)\left[E-{\cal U}(x,\eta)\right]^{1/2}\}^{-1}
W_{1}(E,\eta)
\label{65}
\eq
where
\bq
{\Phi}(E)=\int dx\left[E-{\cal U}(x,{\eta})\right]^{1/2} \label{65b}
\eq
and  a prime now denotes a ${\partial}/{\partial}E$ derivative. Here the
integration extends over the values of $x$ along the unperturbed
orbit associated with $E$. Note that the prefactor of $W_{1}$ is simply
the relative amount of time that, for fixed energy, the system spends
at each point $x$.

By integrating the Fokker-Planck equation for $W(x,E,\eta)$ over $x$, one
obtains the desired equation for $W_{1}(E,{\eta})$, which takes the
form
\ba
&&
{\pa\over\pa \eta}W_{1}(E,{\eta})=\nonumber\\
&&-\left({\pa\over\pa
E}\left[{1\over\chi'(E)}\left\{{\lambda}_{0}\left({\Phi}(E)-
k_BT{\Phi}'(E)\right) +
2{\lambda}_{1}\left({\chi}(E)-k_BT{\chi}'(E)\right)\right.\right.\right.
\nonumber\\
&&\left.\left.\left.+{\lambda}_{2}\left({\Psi}(E)-
k_BT{\Psi}'(E)\right)\right\}\right]\right.\nonumber\\
&&\left. +k_{B}T{{\partial}^{2}\over
{\partial}E^{2}}\left[{1\over\chi'(E)}\left\{{\lambda}_{0}{\Phi}(E) +
2{\lambda}_{1}{\chi}(E) +{\lambda}_{2}{\Psi}(E)\right\}\right]\right)
W_{1}(E,{\eta}), \label{66}
\ea
where
\bq
{\chi}(E)=\int {\,}dx{\,}x{\,}\left[E-{\cal U}(x,{\eta})\right]^{1/2}
\label{67}
\eq
and
\bq
{\Psi}(E)=\int {\,}dx{\,}x^{2}{\,}\left[E-{\cal U}(x,{\eta})\right]^{1/2}.
\label{68}
\eq

Unfortunately, for a generic potential ${\cal U}$ the functions
${\Phi}(E)$, ${\chi}(E)$, and ${\Psi}(E)$ cannot be evaluated
analytically, so that one cannot realize the right hand side of
(\ref{68}) explicitly in terms of simple functions of $E$. For a
quartic potential, for example, these functions can only be
expressed as elliptic integrals, which must be evaluated numerically.
In the limit when the energy $E$ is small and the system is
oscillating about a local minimum of ${\cal U}$ one can proceed
analytically by evaluating the orbit integrals, assuming
that the system is effectively evolving in a simple harmonic
oscillator potential. By doing so, it is straightforward to derive from
the Fokker-Planck equation a transport equation involving the time
derivative of the first energy moment
\bq
\VEV{E({\eta})}{\equiv}\int dE~E~W_{1}(E,{\eta}). \label{69}
\eq

First suppose that ${\Sigma}$ is positive. Then the moment equation
takes the form
\bq
{\pa\over\pa\eta}\VEV{E({\eta})} =
k_{B}T{\lambda}_{0}-
\left({\lambda}_{0}-{k_{B}T{\lambda}_{2}\over
{\omega}_{0}^{~2}}\right)\VEV{E({\eta})}-
{{\lambda}_{2}\over {\omega}_{0}^{~2}}
\VEV{E^{2}({\eta})}. \label{70}
\eq
Note that, because of the reflection symmetry $x \to -x$ for the
potential ${\cal U}$, the functions ${\chi}={\chi}'=0$, so that the
contributions involving ${\lambda}_{1}$ vanish identically.

Unfortunately, this equation still cannot be solved exactly for
$\VEV{E({\eta})}$, as it involves the unknown function
$\VEV{E^{2}({\eta})}$. To obtain a formula for $\VEV{E^{2}({\eta})}$,
one must consider the second moment equation, which in turn relates
$d\VEV{E^{2}({\eta})}/d{\eta}$ to the third moment
$\VEV{E^{3}({\eta})}$. In the spirit of (say) the BBGKY hierarchy, one
requires a truncation approximation.

As in Refs. \cite{RLS} and \cite{LS}, suppose that
\bq
\VEV{E^{2}(t)}~\approx~\kappa\VEV{E(t)}^{2}, \label{71}
\eq
with ${\kappa}=2$. One knows that, when the system is at equilibrium,
with energy $E=k_{B}T$, this equation is satisfied identically for
${\kappa}=2$, and one might expect on physical grounds that, before the
system is ``at equilibrium,'' the energy distribution will be narrower
and ${\kappa}<2$. As emphasized by Lindenberg and Seshadri
\cite{LS}, this truncation approximation thus leads to an upper
limit on the time scale on which the system ``equilibrates'' with the
bath. Given this truncation, one can immediately write down the
solution \cite{LS}
\bq
\VEV{E({\eta})}=
{k_{B}T\left(E_{0}+Ak_{B}T\right) -
AkT\left(k_{B}T-E_{0}\right)\exp\left\{
-\left[(A+1)/A\right]{\lambda}_{0}{\eta} \right\}
\over \left(E_{0}+Ak_{B}T\right)-\left(k_{B}T-E_{0}\right)\exp\left\{
-\left[(A+1)/A\right]{\lambda}_{0}{\eta} \right\} } ,
\label{72}
\eq
where
\bq
A \equiv {\l_0\omega_0^{~2}\over{\lambda}_{2}k_{B}T} .
\label{73}
\eq

In the limit that ${\lambda}_{2} \to 0$, the system approaches an
``equilibrium'' with $\VEV{E}=k_{B}T$ on a time scale $t_{R} \sim
\lambda_0^{-1}$. If ${\lambda}_{2} \neq 0$, the system still evolves
towards an equilibrium with $\VEV{E}=k_{B}T$, but the time scale
$t_{R}$ can be quite different. In fact, in the limit that
${\lambda}_{0} \to 0$,
\bq
t_{R}\sim\omega_0^{~2}/({\lambda}_{2}k_{B}T).  \label{nrel}
\eq
Therefore, when the nonlinear coupling is sufficiently strong, the
system may be driven towards equilibrium, {\em not} by the ordinary
additive noise associated with the linear coupling, but primarily by the
multiplicative noise associated with the nonlinear coupling.

The only point that remains to be checked is that one is
still assuming, as is implicit in this ``envelope'' approximation,
that the time scale ${\omega}_{0}^{-1}$ is much shorter than the
damping time. This, however, is clearly the case when ${\lambda}_{0}$
and ${\lambda}_{2}$ are not too large.  Indeed, one verifies that (a)
the weak damping approximation {\em is} legitimate but (b)
multiplicative noise dominates the evolution towards an equilibrium
whenever \cite{LS}
\bq
{{\lambda}_{0}\over {\omega}_{0}}{\;}{\ll}{\;}
{{\lambda}_{0}{\omega}_{0}^{~2}\over {\lambda}_{2}k_{B}T}
{\;}{\ll}{\;} 1 .  \label{74}
\eq

Consider now the case ${\Sigma}<0$ and the system oscillating about
one of the two minima of the potential $x_{0}{\ne}0$. (This is the
case relevant to first order phase transitions.) Here the terms
involving ${\chi}$ and its energy  derivative do not vanish, and the
formulae for ${\Phi}$ and ${\Psi}$ acquire additional terms involving
the location $x_{0}$ of the new minimum. However, one still recovers a
relatively simple exact equation for
${\partial}\VEV{E}/{\partial}{\eta}$. In this case, the moment
equation is
\bq
{\pa\over\pa\eta}\VEV{E({\eta})}=
k_{B}TL-\left(L-{k_{B}T{\lambda}_{2}\over
{\omega}_{0}^{~2}}\right)\VEV{{\cal E}({\eta})}-
{{\lambda}_{2}\over 2{\omega}_{0}^{~2}}
\VEV{{\cal E}^{2}({\eta})}, \label{75}
\eq
where
\bq
{\cal E}=E-{\cal U}(x_{0}) \label{76}
\eq
denotes the system energy defined relative to the minimum of the
potential, and
\bq
L={\lambda}_{0}+{\lambda}_{1}x_{0}+{\lambda}_{0}x_{0}^{~2} \label{77}
\eq
plays the role of a ``dressed'' ${\lambda}_{0}$. Note in particular
that, if $\abs{x_{0}}$ is large, as will be the case when $\Sigma \ll
\Lambda$, the dressed $L$ can be much larger than ${\lambda}_{0}$.
This implies that, in this case, the nonlinear couplings can reduce
the overall equilibration time {\em both} through the introduction of
a new term $\propto \l_{2}\VEV{E^{2}}$ {\em and} through an increase
in the effective linear coupling.

In any case, as long as ${\cal U}(x_{0},{\eta})$ is only slowly
varying in time, the derivative
${{\partial}\VEV{E({\eta})}/{\partial}{\eta}}$ can be replaced by
${{\partial}\VEV{{\cal E}({\eta})}/{\partial}{\eta}}$. And, to the
extent that the coefficients $L$ and ${\lambda}_{2}$ may be
approximated as time-independent, (\ref{63}) can be solved
analytically. The result is an expression identical to (\ref{56}),
except that $E$ is replaced by the shifted ${\cal E}=E-{\cal U}_{0}$
and
\bq
A={L{\omega}_{0}^{~2}\over{\lambda}_{2}k_{B}T}. \label{78}
\eq

Numerical tests of these results are not difficult to carry out. A
simple test is to take a harmonic oscillator and couple it
quadratically to noise. Simulations show, that in the weak coupling
regime, the relaxation to equilibrium is indeed much faster than for
the case of linear coupling and that the relaxation time scale is in
agreement with (\ref{nrel}) \cite{SHNN}. In the case of strong
coupling to the heat bath, the energy-envelope method is no longer
applicable. Numerical simulations indicate that the relaxation time
scales for both the additive and multiplicative noises are similar at
large coupling.

An interesting example that corresponds to the case $\Sigma < 0$
occurs in the problem of determining the nucleation rate of kinks in a
double-well $\f^4$ theory in $1+1$-dimensions. One begins with the
system in one vacuum and counts kinks as they nucleate (as pieces of
the field configuration ``hop'' to the other side). At low
temperatures compared to the kink energy, the nucleation rate is very
low (and is a significant problem when carrying out numerical
simulations). However, in the presence of multiplicative noise the
rate can be significantly enhanced by  following an argument exactly
analogous to the one given here for the particle system. While the
theory has not yet been completely worked out, numerical simulations
do show a significant increase in the nucleation rate in the case of
nonlinear noise \cite{SHNN}.

\newpage
\centerline{\bf IV. Conclusion}

The objective here has been to motivate and derive exact Langevin
equations for particle systems and field theories using an oscillator
model for the heat bath. Multiplicative noise is seen to arise in
these models as a consequence of nonlinear coupling to the heat bath,
a situation particularly natural when dealing with systems with
constraints. A particularly interesting consequence of nonlinear noise
is the drastic change possible in relaxation time scales: a simple
analytic estimate is possible in the case of weak damping using the
energy-envelope method. Numerical simulations have been carried out
to confirm the theoretical prediction. While not yet complete, they
are in agreement with the theory so far.

There are several interesting applications of this approach.
Langevin simulations for determining the rate of sphaleron transitions
at finite temperature in the nonlinear sigma model have recently been
carried out using the nonlinear Langevin equation derived here. Such
simulations are also easy to carry out for spin chains using
essentially the same technology. The problem of the dependence of the
nucleation rate of kinks on the nature of the noise (additive or
multiplicative) is another interesting issue. There is clear numerical
evidence that the nucleation rate is speeded up by multiplicative
noise but a complete analytic theory has still to be worked out. Work
in this direction is in progress.

One of the original motivations for this work was to study the onset
of inflation in such a way as to answer the question first raised by
Mazenko, Unruh, and Wald \cite{MUW}: Does a homogeneous slow-roll
really take place or does domain formation happen so quickly that
inflation is ruled out? This question was tackled by Albrecht and
Brandenberger \cite{AB} who argued that for a sufficiently small value
of the self-coupling for the inflaton field the scenario remains
viable. Recently, the effect of fluctuations was studied in Ref. \cite
{RB} via an essentially {\em ad hoc} approach, the results of which
did not weaken the estimates of Ref. \cite {AB}. Now, when one deals
with an inflaton field with a (small) quartic self-coupling, the
effective equations for the mean field contain a (weak) nonlinear
damping term. It is then reasonable to ask what happens when the
relaxation to the true minimum is controlled by these terms. The work
presented here suggests that these terms can dramatically change this
time scale. Whether the change is of practical relevance can only be
determined by further calculations, both analytic and numerical.

\centerline{\bf Acknowledgements}

I would like to thank the organizers, J. Robert Buchler, George
Contopoulos, and Henry E. Kandrup for a very enjoyable and
thought-provoking meeting. I am indebted to my collaborators Frank
Alexander, Emil Mottola, and especially, Henry Kandrup for many
discussions of the ideas presented here. I would also like to thank J.
Robert Dorfman and Robert Zwanzig for several inspiring discussions.
Support for this research was provided by the Department of Energy and
the Air Force Office for Scientific Research.


\newpage
\begin{thebibliography}{99}
\bibitem{PW} G. Parisi and Y. S. Wu, Sci. Sin. {\bf 24}, 483 (1981).
For a good review, see, P. Damgaard and H. Huffel, Phys. Rep. {bf
152}, 227 (1987).
\bibitem{HL} W. Horsthemke and R. Lefever, {\em Noise Induced
Transitions} (Springer-Verlag, New York, 1984).
\bibitem{AAS} A. A. Starobinsky, in {\em Fundamental Interactions}
(MGPI Press, Moscow, 1983); in {\em Current Topics in Field Theory,
Quantum Gravity, and Strings} edited by H. J. deVega and N. Sanchez
(Springer, New York, 1986). See also, A. Vilenkin, Nuc. Phys. B {\bf
226}, 527 (1983).
\bibitem{QD} A recent review is that of W. H. Zurek, Phys. Today {\bf
44}, 36 (1991).
\bibitem{RB} Cf., R. H. Brandenberger, H. A. Feldman, and J. H.
MacGibbon, Phys. Rev. D {\bf 37}, 2071 (1988); H. A. Feldman, Phys.
Rev. D {\bf 38}, 459 (1988); R. H. Brandenberger and H. A. Feldman,
Physica A {\bf 158}, 343 (1989).
\bibitem{CB} J. M. Cornwall and R. Bruinsma, Phys. Rev. D {\bf 38},
3146 (1988).
\bibitem{HKA} S. Habib and H. E. Kandrup, Phys. Rev. D {\bf 39}, 2871
(1989).
\bibitem{HKB} S. Habib and H. E. Kandrup, Phys. Rev. D {\bf 46}, 5303
(1992).
\bibitem{DECO} Cf., S. Habib and R. Laflamme, Phys. Rev. D {\bf 42},
4056 (1990); J. P. Paz and S. Sinha, Phys. Rev. D {\bf 44}, 1038
(1991); and references therein.
\bibitem{BLH} Cf., E. Calzetta and B. L. Hu, Phys. Rev. D {\bf 40},
656 (1989).
\bibitem{SHSI} S. Habib, Phys. Rev. D {\bf 46}, 2408 (1992).
\bibitem{RWZ} R. Zwanzig, J. Stat. Phys. {\bf 9}, 215 (1973).
\bibitem{LS} K. Lindenberg and V. Seshadri, Physica {\bf 109A}, 481
(1981).
\bibitem{RLS} R. L. Stratonovich, {\em Topics in the Theory of Random
Noise} (Gordon and Breach, New York, 1967), Vol. 1, p. 115.
\bibitem{MS} K. M\"{o}hring and U. Smilansky, Nucl. Phys. {\bf A338},
227 (1980).
\bibitem{CLA} A. O. Caldeira and A. J. Leggett, Physica {\bf 121A},
587 (1983); Ann. Phys. (N.Y.) {\bf 149}, 374 (1983).
\bibitem{IO} This model appears in many guises in the literature. For
a discussion, see G. W. Ford, J. T. Lewis, and R. F. O'Connell, Phys.
Rev. A {\bf 37}, 4419 (1988).
\bibitem{FJSH} See, e.g., F. J. Alexander and S. Habib, Phys. Rev.
Lett. {\bf 71}, 955 (1993). (Earlier work is
referenced and discussed in this paper.)
\bibitem{NTS} S. Habib and E. Mottola, (in preparation).
\bibitem{SHNN} S. Habib, (in preparation).
\bibitem{MUW} G. Mazenko, W. G. Unruh, and R. M. Wald, Phys. Rev. D
{\bf 31}, 273 (1985).
\bibitem{AB} A. Albrecht and R. H. Brandenberger, Phys. Rev. D {\bf
31}, 1225 (1985).
\end{thebibliography}
\end{document}